\documentclass[onecollarge,natbib]{svjour2}
\bibpunct{[}{]}{,}{n}{}{,} 

\smartqed  

\usepackage{graphicx}
\usepackage{mathptmx}      
\usepackage{latexsym}
\journalname{Few-Body Systems} 

\newcommand{\eq}{\begin{eqnarray}}
\newcommand{\en}{\end{eqnarray}}

\begin{document}

\title{Mesons and baryons in a soft-wall holographic approach 
\thanks{This work was supported by Federal Targeted Program "Scientific
and scientific-pedagogical personnel of innovative Russia"
Contract No. 02.740.11.0238, by FONDECYT (Chile) under Grant No. 1100287. \\
A.V. acknowledges the financial support from FONDECYT (Chile) 
Grant No. 3100028.} 
\thanks{Presented by Valery E. Lyubovitskij at LIGHTCONE 2011, 23 - 27 May, 
2011, Dallas.}
}
\titlerunning{Mesons and baryons in a soft-wall holographic approach} 

\author{Thomas~Gutsche$^1$,~Valery~E.~Lyubovitskij$^{1,\ast}$\thanks{$\ast$ 
On leave of absence from Department of Physics, Tomsk State University, 
634050 Tomsk, Russia},~Ivan~Schmidt$^2${,\,}Alfredo~Vega$^2$}

\authorrunning{Thomas Gutsche, Valery E. Lyubovitskij, 
Ivan Schmidt, Alfredo Vega} 
   
\institute{$^1$ Institut f\"ur Theoretische Physik,
                Universit\"at T\"ubingen, 
                Kepler Center for Astro and Particle Physics, \\
                Auf der Morgenstelle 14, D-72076 T\"ubingen, Germany
                \and  \\ 
           $^2$ Departamento de F\'\i sica y Centro Cient\'\i fico 
                Tecnol\'ogico de Valpara\'\i so (CCTVal), 
                Universidad T\'ecnica Federico Santa Mar\'\i a, 
                Casilla 110-V, Valpara\'\i so, Chile
}

\date{Received: date / Accepted: date}

\maketitle

\begin{abstract} 
We discuss a holographic soft-wall model developed for the
description of mesons and baryons with adjustable quantum numbers 
$n, J, L, S$. This approach is based on an action which describes
hadrons with broken conformal invariance and which incorporates
confinement through the presence of a background dilaton field. 

\keywords{Soft-wall holographic model, dilaton, hadrons, mass spectrum} 
\end{abstract}

\section{Introduction}
\label{intro}

Based on the correspondence of string theory in anti-de Sitter (AdS)
space and conformal field theory (CFT) in physical
space-time~\cite{Maldacena:1997re}, a class of AdS/QCD approaches
was recently successfully developed for describing the phenomenology
of hadronic properties. In order to break conformal invariance and
incorporate confinement in the infrared (IR) region two alternative
AdS/QCD backgrounds have been suggested in the literature: the
``hard-wall'' approach~\cite{Hard_wall1}, based on
the introduction of an IR brane cutoff in the fifth dimension, and
the ``soft-wall'' approach~\cite{Soft_wall1}, based on
using a soft cutoff. This last procedure can be introduced in the
following ways: i) as a background field (dilaton) in the overall
exponential of the action, ii) in the warping factor of the AdS
metric, iii) in the effective potential of the action. These methods
are in principle equivalent to each other due to a redefinition of
the bulk field involving the dilaton field or by a
redefinition of the effective potential.  
In the literature there exist detailed discussions of the sign
of the dilaton profile in the dilaton exponential
$\exp(\pm\varphi)$~\cite{Soft_wall1,Soft_wall2a,Soft_wall2b,%
SWminus,SWplus} for the soft-wall model (for a discussion of the
sign of the dilaton in the warping factor of the AdS metric see
Refs.~\cite{Soft_wall3a}). The negative sign was suggested in
Ref.~\cite{Soft_wall1} and recently discussed in Ref.~\cite{SWplus}.
It leads to a Regge-like behavior of the meson spectrum, including a
straightforward extension to fields of higher spin~$J$. Also, in
Ref.~\cite{SWplus} it was shown that this choice of the dilaton sign
guarantees the absence of a spurious massless scalar mode in the
vector channel of the soft-wall model. We stress that alternative
versions of this model with a positive sign are also possible. One
should just redefine the bulk field $V$ as $V=\exp(\varphi) \tilde V$, 
where the transformed field corresponds to the dilaton with an opposite 
profile. It is clear that the underlying action changes, and extra 
potential terms are generated depending on the dilaton field 
(see detailed discussion in~\cite{Gutsche:2011vb}). 

\section{Bosonic case}
\label{sec:1}

First we demonstrate the equivalence of both versions of the
soft-wall model with a positive and negative dilaton profile in the
case of the bosonic field, and afterwards we consider the fermionic
field and extension to higher values of total angular momentum $J$.
We consider the propagation of a scalar field $S(x,z)$ in $d+1$
dimensional AdS space. The AdS metric is specified by 
$
ds^2 =
g_{MN} dx^M dx^N = \eta_{ab} \, e^{2A(z)} \, dx^a dx^b = e^{2A(z)}
\, (\eta_{\mu\nu} dx^\mu dx^\nu - dz^2)\,, \ 
\eta_{\mu\nu} = {\rm diag}(1, -1, \ldots, -1)$, 
where $M$ and
$N = 0, 1, \cdots , d$ are the space-time (base manifold) indices,
$a=(\mu,z)$ and $b=(\nu,z)$ are the local Lorentz (tangent) indices,
$g_{MN}$ and  $\eta_{ab}$ are curved and flat metric tensors, which
are related by the vielbein $\epsilon_M^a(z)= e^{A(z)} \,
\delta_M^a$ as $g_{MN} =\epsilon_M^a \epsilon_N^b \eta_{ab}$. Here
$z$ is the holographic coordinate, $R$ is the AdS radius, and $g =
|{\rm det} g_{MN}| = e^{2 A(z) (d+1)}$. In the following we restrict
ourselves to a conformal-invariant metric with $A(z) = \log(R/z)$.

The actions for the scalar field $(J=0)$ with
a positive and negative dilaton are~\cite{Soft_wall2a,Soft_wall2b}
\eq
S^+_0 = \frac{1}{2} \int d^dx dz \, \sqrt{g} \, e^{\varphi(z)} \,
\biggl[ g^{MN} \partial_M S^+(x,z) \partial_N S^+(x,z)
- \mu_0^2 \, S^+(x,z)S^+(x,z) \biggr]
\en
and~\cite{Soft_wall8}
\eq
S^-_0 = \frac{1}{2} \int d^dx dz \sqrt{g} e^{-\varphi(z)}
\biggl[ g^{MN} \partial_M S^-(x,z) \partial_N S^-(x,z)
- \Big(\mu_0^2 + \Delta V_0(z)\Big) \, S^-(x,z)S^-(x,z) \biggr] \,.
\en
The superscripts $+$ and $-$ correspond to the cases of 
a positive or negative dilaton, respectively.
The actions are equivalent to each other, which is obvious by the
bulk field redefinition: 
$S^\pm(x,z) = e^{\mp\varphi(z)} S^\mp(x,z)$. 
The difference between the two actions is absorbed in the effective
potential $\Delta V_0(z) = e^{-2A(z)} \Delta U_0(z)$,
where 
$\Delta U_0(z) = \varphi^{\prime\prime}(z)
+ (d-1) \, \varphi^{\prime}(z) A^{\prime}(z)$. 
The quantity $\mu_0^2 R^{2}= \Delta (\Delta - d)$ is
the bulk boson mass, where $\Delta$ is the dimension of the
interpolating operator dual to the scalar bulk field.
For the case of the bulk fields dual to the scalar fields
$\Delta = 2 + L$, where $L = {\rm max} \, | L_z |$ is the maximal value of the
$z$-component of the orbital angular momentum~\cite{Soft_wall2a,Soft_wall2b}. 
In particular, we have $L=0$ for $J^P = 0^-$ states and 
$L=1$ for $J^P = 0^+$ states. Notice that $\Delta$ is identified
with the twist $\tau$ of two-parton states. Later we will show that $\tau$
for meson states is independent on the total angular momentum $J$,  i.e.
$\Delta_J \equiv \tau = 2 + L$. Notice that both actions have the correct
conformal limit at $z \to 0$, where the dilaton field vanishes and
conformal invariance is restored.

In a next step we modify the above form of the actions to obtain expressions
which are more convenient in applications. First, one can remove the
dilaton field from the overall exponential by a specific redefinition of the
bulk field $S^\pm$ as: $S^\pm(x,z) = e^{\mp\varphi(z)/2} S(x,z)$. 
In terms of the field $S(x,z)$ the transformed action,
which now is universal for both versions of the soft-wall model,
reads~\cite{Soft_wall2a,Soft_wall2b}:
\eq\label{S_action}
S_0 = \frac{1}{2} \int d^dx dz \sqrt{g} \,
\biggl[ g^{MN} \partial_M S(x,z) \partial_N S(x,z)
- (\mu_0^2 + V_0(z)) S^2(x,z)  \biggr] \,,
\en
where $V_0(z) = e^{-2A(z)} U_0(z)$ and with the effective potential
\eq
U_0(z) = \frac{1}{2} \varphi^{\prime\prime}(z)
+ \frac{1}{4} (\varphi^{\prime}(z))^2
+ \frac{d-1}{2}\varphi^{\prime}(z) A^{\prime}(z) \,.
\en
The last expression is identical with
the light-front effective potential found in Ref.~\cite{Soft_wall2b}
for $d=4$, $J=0$ [see Eq.(10)]. 
With Lorentzian signature the action~(\ref{S_action}) is given by
\eq\label{S_hidden}
S_0 &=&  \frac{1}{2} \int d^dx dz \, e^{B_0(z)}
\biggl[ \partial_\mu S(x,z) \partial^\mu S(x,z)
- \partial_z S(x,z) \partial_z S(x,z)
- \Big(e^{2A(z)}\mu^2 + U_0(z)\Big) S^2(x,z) \biggr] \,,\nonumber\\
B_0(z) &=& (d-1) \, A(z) \,.
\en
Then we use a Kaluza-Klein (KK) expansion  
$S(x,z) = \sum_n \ S_n(x) \ \Phi_{n}(z)$, 
where $n$ is the radial quantum number, $S_n(x)$
is the tower of the KK modes dual to
scalar mesons and $\Phi_{n}$ are their extra-dimensional profiles
(wave-functions). We suppose a free propagation
of the bulk field along the $d$ Poincar\'e coordinates with four-momentum
$p$, and a constrained propagation along the $(d+1)$-th coordinate $z$
(due to confinement imposed by the dilaton field). 

Performing the substitution $\Phi_{n}(z) = e^{-B_0(z)/2} \phi_{n}(z)$ 
and restricting to mass-shell $p^2 = M_{n0}^2$ 
we derive the Schr\"odinger-type EOM for $\phi_{n}(z)$:
\eq\label{Eq1}
\Big[ - \frac{d^2}{dz^2} + \frac{4L^2 - 1}{4z^2} + U_0(z)
\Big] \phi_{n}(z) = M^2_{n0} \phi_{n}(z)
\en
where
\eq
\phi_{n}(z) = \sqrt{\frac{2 \Gamma(n+1)}{\Gamma(n+L+1)}} \ \kappa^{L+1}
\ z^{L+1/2} \ e^{-\kappa^2 z^2/2} \ L_n^{L}(\kappa^2z^2)
\en
and 
$M^2_{n0} = 4 \kappa^2 \Big( n + \frac{L}{2} \Big)$ 
is the mass spectrum of scalar field.
Here we use the generalized Laguerre polynomials $L_n^m(x)$. 
Notice that the normalizable mode $\Phi_{n}(z)$ 
has the correct behavior in both the 
ultraviolet (UV) and infrared (IR) limits:
$\Phi_{n}(z) \ \sim \ z^{2+L}
\ \ {\rm at \ small} \ z\,, \quad 
\Phi_{n}(z) \to 0 \ \ {\rm at \ large} \ z$. 
Using the KK expansion, EOM for the KK profiles
$\Phi_{n}(z)$, the $d+1$-dimensional action for the bulk field reduces
to a $d$-dimensional action for the scalar fields $S_n(x)$
dual to scalar mesons with masses $M_{n0}$:
\eq
S_0^{(d)} = \frac{1}{2} \sum\limits_n \, \int d^dx
\biggl[ \partial_\mu S_n(x) \, \partial^\mu S_n(x)
- M_{n0}^2 S_n^2(x) \biggr] \,. 
\en
This last equation is a manifestation of the gauge-gravity duality.
In particular, it explicitly demonstrates
that effective actions for conventional hadrons in $d$ dimensions
can be generated from actions for bulk
fields propagating in extra $d+1$ dimensions. The impact of the 
extra-dimension is encoded in the hadronic mass squared $M_n^2$, 
which is the solution of the Schr\"odinger equation~(\ref{Eq1}) 
for the KK profile in extra dimension $\phi_{n}(z)$. 
For the case of the vector field $V_M(x,z)$ 
we proceed by analogy. 
After straightforward algebra we derive the Schr\"odinger-type EOM for
the KK mode and find the mass spectrum of vector mesons 
$M^2_{n1} = 4 \kappa^2 ( n + \frac{L}{2}+\frac{1}{2} )$. 

We further consider bulk boson fields with higher values of
$J \ge 2$. This problem, in the context of soft-wall models, has
been considered before in
Refs.~\cite{Soft_wall1,Soft_wall2a,Soft_wall8,Soft_wall3b,Soft_wall5}. 
In particular, it was shown that the soft-wall model
reproduces the Regge-behavior of the mesonic mass spectrum
$M_{nJ}^2 \sim n + J$. Here, extending our results for scalar and
vector fields, we show that the bound-state problem is independent on
the sign of the dilaton profile.

We describe a bosonic spin-$J$ field $\Phi_{M_1 \cdots M_J}(x,z)$
by a symmetric, traceless tensor, satisfying the conditions 
$\nabla^{M_1}  \Phi_{M_1M_2 \cdots M_J} = 0\,, \ 
g^{M_1M_2}  \Phi_{M_1M_2 \cdots M_J} = 0$. 
The actions for the bulk field $\Phi_J$ with positive and negative
dilatons are~\cite{Soft_wall2a,Soft_wall2b}
\eq\label{SW_JPlus_Minus}
S^\pm_J &=& \frac{(-)^J}{2}
\int d^dx dz \, \sqrt{g} \, e^{\pm\varphi(z)} \,
\biggl[ g^{MN} g^{M_1N_1} \cdots g^{M_JN_J}
\nabla_M\Phi_{M_1 \cdots M_J}^\pm(x,z)
\nabla_N\Phi_{N_1 \cdots N_J}^\pm(x,z) \nonumber\\
&-& \Big(\mu_J^2 + \Delta V_J^\pm(z)\Big)  \, g^{M_1N_1} \cdots g^{M_JN_J}
\Phi_{M_1 \cdots M_J}^\pm(x,z)\Phi_{N_1 \cdots N_J}^\pm(x,z) \biggr]  \,, 
\en 
where $V_J^+ = 0$ and $V_J^- = V_J$. 
Here $\nabla_M$ is the covariant derivative
with respect to AdS coordinates.  

In Refs.~\cite{Soft_wall2a,Soft_wall2b,Soft_wall8}
higher spin fields have been considered in a ``weak gravity'' approximation,
restricting the analysis to flat metric and therefore
identifying the covariant derivative with the normal derivative
(i.e. neglecting the affine connection).
First, we review these results and then consider the
general case with covariant derivatives. In the following we
call the scenario with normal derivatives scenario I
and the scenario with covariant derivatives scenario II.

In scenario I the bulk mass is given by
by~\cite{Soft_wall2a,Soft_wall2b,Soft_wall8} 
$\mu_J^2 R^2 = (\Delta - J) (\Delta + J - d)$, 
which is fixed by the behavior of bulk fields $\Phi_J$ near the ultraviolet
boundary $z=0$. The potential  $\Delta V_J(z) = e^{-2A(z)} \Delta U_J(z)$
is given by
$\Delta U_J(z) =  \varphi^{\prime\prime}(z)
+ (d-1-2J) \, \varphi^\prime(z) \, A^\prime(z)$. 
Notice that both quantities $\mu_J^2$ and $\Delta U_J(z)$
are generalizations of the scalar ($J=0$) and vector ($J=1$) cases
considered before. In particular, they are related to
those for the scalar field as follows:
$\mu_J^2 R^2  = \mu_0^2 R^2 + J (d - J)$, 
$\Delta U_J(z) = \Delta U_0(z) - 2J \varphi^\prime(z) A^\prime(z)$. 
As before the two actions can be reduced to the action with 
a dilaton hidden in an additional potential term, using the 
transformation $\Phi_J^{\pm}(x,z) = e^{\mp\varphi(z)/2} \Phi_J(x,z)$. 
Then the action takes the form 
\eq
S_J &=& \frac{(-)^J}{2}
\int d^d x dz \sqrt{g}
\biggl[ g^{MN} g^{M_1N_1} \cdots g^{M_JN_J} \,
\partial_M\Phi_{M_1 \cdots M_J}^+(x,z) \,
\partial_N\Phi_{N_1 \cdots N_J}^+(x,z) \nonumber\\
&-& (\mu_J^2 + V_J(z)) \, g^{M_1N_1} \cdots g^{M_JN_J}
\Phi_{M_1 \cdots M_J}^+(x,z) \, \Phi_{N_1 \cdots N_J}^+(x,z) \biggr]
\en
where $V_J(z) = e^{-2A(z)} U_J(z)$, and with the effective potential
\eq
U_J(z) = \frac{1}{2} \varphi^{\prime\prime}(z)
+ \frac{1}{4} (\varphi^{\prime}(z))^2
+ \frac{d-1-2J}{2}\varphi^{\prime}(z) A^{\prime}(z) \,.
\en
This last expression is identical with
the light-front effective potential found in Ref.~\cite{Soft_wall2b}
for $d=4$ and arbitrary $J$ [see Eq.(10)]:
$U_J(z) = \kappa^4 z^2 + 2 \kappa^2 (J - 1)$. 
Using standard algebra and restricting to the axial gauge
$\Phi_{\cdots z \cdots}(x,z)=0$, writing down the action
in terms of fields with Lorentz indices 
and rescaling the fields by the
boost (total angular momentum) factor  $e^{J A(z)}$ as
$\Phi_{\mu_1 \cdots \mu_J}(x,z) \to  e^{J A(z)} \,
 \Phi_{\mu_1 \cdots \mu_J}(x,z)$ 
we write down the action in the form 
\eq\label{Lorentz_actionPhiJ2}
\hspace*{-.5cm}S_J &=& \frac{(-)^J}{2} \int d^d x dz \, e^{B_0(z)}
\biggl[
  \partial_\mu\Phi_{\mu_1 \cdots \mu_J}(x,z)
  \partial^\mu\Phi_{\mu_1 \cdots \mu_J}(x,z)
- \partial_z\Phi_{\mu_1 \cdots \mu_J}(x,z)
  \partial_z\Phi_{\mu_1 \cdots \mu_J}(x,z) \nonumber\\
&-& \Big(e^{2A(z)} \mu_0^2 + U_J(z)\Big) \Phi_{\mu_1 \cdots \mu_J}(x,z)
    \Phi^{\mu_1 \cdots \mu_J}(x,z) \biggr] \,.
\en
Doing the KK expansion
$\Phi^{\mu_1 \cdots \mu_J}(x,z)
= \sum_n \ \Phi^{\mu_1 \cdots \mu_J}_n(x) \ \Phi_{n}(z)$ 
and the substitution 
$\Phi_{n}(z) = e^{-B_0(z)/2} \phi_{n}(z)$ 
we derive the Schr\"odinger-type EOM for
$\phi_{n}(z)$:
\eq\label{Eq1J}
\Big[ - \frac{d^2}{dz^2} + \frac{4L^2 - 1}{4z^2} + U_J(z)
\Big] \phi_{n}(z) = M^2_{nJ} \phi_{n}(z),
\en
where 
$M^2_{nJ} = 4 \kappa^2 ( n + \frac{L}{2} + \frac{J}{2} )$ 
is the mass spectrum of higher $J$ fields, which generalizes our results
for scalar and vector fields. 
At large values of $J$ or $L$ we reproduce the Regge behavior
of the meson mass spectrum:  $M^2_{nJ} \sim  n + J$.  
Finally, using the KK expansion and the EOMs for the wave functions,
we derive the $d$-dimensional action for mesons with total angular
momentum $J \ge 2$ and masses $M^2_{nJ}$:
\eq
S_J^{(d)} = \frac{(-)^J}{2} \sum\limits_n \, \int d^dx
\biggl[ \partial_\mu\Phi_{\mu_1 \dots \mu_J, n}(x) \,
\partial^\mu\Phi^{\mu_1 \dots \mu_J}_n(x)
- M_{nJ}^2 \Phi_{\mu_1 \cdots \mu_J, n}(x)
\Phi^{\mu_1 \cdots \mu_J}_n(x) \biggr] \,.
\en 
Now we consider scenario II, i.e. without truncation
of covariant derivatives. The gauge-invariant actions for
the totally symmetric higher spin boson fields have been considered
e.g. in Refs.~\cite{Buchbinder:2001bs}.
In this case the bulk mass is fixed by gauge invariance, and given by
$\mu_J^2 R^2 = J^2 + J (d-5) + 4 - 2d$. 
This mass leads to the following results for the scaling of the KK profiles:
$\Phi_{n}(z) \ \sim \ z^{2+J}$ at $z \to 0$, and their masses
$M^2_{nJ} = 4 \kappa^2 ( n + J )$, which are acceptable only
for the limiting cases $J=L$ and $J\to \infty$.
Notice that the soft-wall actions~(\ref{SW_JPlus_Minus}) 
are obtained from gauge-invariant actions for totally symmetric higher
spin boson fields~\cite{Buchbinder:2001bs}
via the introduction of the dilaton field, which breaks conformal and gauge
invariance. Therefore, it is not necessary to use the bulk mass
constrained by gauge invariance. In particular, in order to get correct
scaling the the KK profile $\Phi_{n}(z) \ \sim \ z^{2+L}$ and their masses
we should use $\mu_J^2 R^2 = ( \Delta - J ) ( \Delta + J - d) - J
= \mu_0^2 R^2 + J (d - 1 - J)$. 
In this case scenario II is fully equivalent to scenario I.

\section{Fermionic case}

In the fermion case we first
consider the low-lying $J=1/2$ modes $\Psi_\pm(x,z)$
(here the index $\pm$ corresponds again to scenarios with
positive/negative dilaton profiles, respectively.)
The actions with positive and negative dilaton
read ~\cite{Soft_wall7,Soft_wall6,Soft_wall9}
\eq
S^\pm_{1/2} =  \int d^dx dz \, \sqrt{g} \, e^{\pm\varphi(z)} \,
\biggl[ \frac{i}{2} \bar\Psi^\pm \epsilon_a^M \Gamma^a
{\cal D}_M \Psi^\pm - \frac{i}{2} ({\cal D}_M\Psi^\pm)^\dagger \Gamma^0
\epsilon_a^M \Gamma^a \Psi^\pm - \bar\Psi^\pm \Big(\mu + V_F(z)\Big)
\Psi^\pm \biggr] \,, 
\en 
where ${\cal D}_M$ is the covariant derivative 
and $\Gamma^a=(\gamma^\mu, - i\gamma^5)$ are the Dirac matrices. 

The quantity $\mu$ is the bulk fermion mass with $m = \mu R = \Delta - d/2$,
where $\Delta$ is the dimension of the baryon interpolating
operator, which is related with the scaling dimension $\tau = 3 + L$
as $\Delta = \tau + 1/2$.  For $J=1/2$ we have two baryon multiplets
$J^P = 1/2^+$ for $L=0$ and $J^P = 1/2^-$ for $L=1$.
$V_F(z) = \varphi(z)/R$ is the dilaton field dependent effective potential.
Its presence is necessary due to the following reason.
When fermionic fields are rescaled 
$\Psi^\pm(x,z) = e^{\mp\varphi(z)/2} \Psi(x,z)$, 
the dilaton field is removed from the overall exponential. 
The form of the potential $V_F(z)$ is constrained in order to
get solutions of the EOMs for fermionic KK modes
of left and right chirality, and the correct asymptotics of the nucleon
electromagnetic form factors at large
$Q^2$~\cite{Soft_wall7,Soft_wall6,Soft_wall9}. 

The action in terms of a field with Lorentz indices is:
\eq\label{S_F2}
S_{1/2} = \int d^dx dz \, \sqrt{g} \,
\bar\Psi(x,z)
\biggl[ i\not\!\partial + \gamma^5\partial_z
+ \frac{d}{2} A^\prime(z) \gamma^5
-  \frac{e^{A(z)}}{R} \Big(m + \varphi(z)\Big)  \biggr] \Psi(x,z),
\en
where the Dirac field satisfies the following
EOM~\cite{Soft_wall7,Soft_wall6,Soft_wall9}: 
$[ \ \!\!{iz}\not\!\partial + \gamma^5z\partial_z
- \frac{d}{2} \gamma^5
-  m - \varphi(z) ] \Psi(x,z) = 0$. 
Based on these solutions the fermionic action should be extended by
an extra term in the ultraviolet boundary (see details in
Refs.~\cite{Soft_wall7,Henningson:1998cd}).
Here we review the derivation of the EOMs for the KK modes
dual to the left- and right-chirality spinors, in the soft-wall
model~\cite{Soft_wall7,Soft_wall6,Soft_wall9}.
First we expand the fermion field in left- and right-chirality
components: 
$\Psi(x,z) = \Psi_L(x,z) + \Psi_R(x,z)$. 
Then we perform a KK expansion for the $\Psi_{L/R}(x,z)$ fields: 
$\Psi_{L/R}(x,z) = \sum_n \ \Psi_{L/R}^n(x) \ F^n_{L/R}(z)$. 
The KK modes $F^n_{L/R}(z)$ satisfy the two coupled 
one-dimensional EOMs~\cite{Soft_wall7,Soft_wall6,Soft_wall9}:
\eq
\biggl[\partial_z \pm \frac{e^{A}}{R} \, \Big(m+\varphi\Big)
+ \frac{d}{2} A^\prime \biggr] F^n_{L/R}(z) = \pm M_n
F^n_{R/L}(z) \,,
\en
where $M_n$ is the mass of baryons with $J=1/2$. 
After the substitution 
$F^n_{L/R}(z) = e^{-  A(z) \cdot d/2} \, f^n_{L/R}(z)$ 
we derive decoupled Schr\"odinger-type EOM for $f^n_{L/R}(z)$
\eq
\biggl[ -\partial_z^2
+ \kappa^4 z^2 + 2 \kappa^2 \Big(m \mp \frac{1}{2} \Big)
+ \frac{m (m \pm 1)}{z^2} \biggr] f^n_{L/R}(z) = M_n^2 \, f^n_{L/R}(z),
\en
where 
\eq 
f^n_{L}(z) = c_L 
\ \kappa^{L+3}
\ z^{L+5/2} \ e^{-\kappa^2 z^2/2} \ L_n^{L+2}(\kappa^2z^2)\,, \ \ \ \  
f^n_{R}(z) = c_R  
\ \kappa^{L+2}
\ z^{L+3/2} \ e^{-\kappa^2 z^2/2} \ L_n^{L+1}(\kappa^2z^2)
\en 
and 
$M_n^2 = 4 \kappa^2 ( n + L + 2 )$ 
with $L = m-3/2 = 0, 1$ for $J=1/2$ fermions. 
Here $c_L = \sqrt{2\Gamma(n+1)/\Gamma(n+L+3)}$ 
and $c_R = \sqrt{2\Gamma(n+1)/\Gamma(n+L+2)}$.   
One can see that
the functions $F^n_{L/R}(z)$ have the correct scaling behavior
for small $z$: 
$F^n_{L}(z) \sim z^{9/2+L}$, $F^n_{R}(z) \sim z^{7/2+L}$ 
and vanish at large $z$ (confinement).
As in the bosonic case, integration over the holographic coordinate $z$
gives a $d$-dimensional action for the fermion field
$\Psi^n(x) = \Psi^n_L(x) + \Psi^n_R(x)$: 
\eq 
S_{1/2}^{(d)} = \sum\limits_n \, \int d^dx
\bar\Psi^n(x) [ i \not\!\partial - M_{n} ] \Psi^n(x) \,. 
\en 
Extension of our formalism for higher spin states $J$
is straightforward. In particular, the actions for higher spin 
fermions with positive and negative dilaton are written as
\eq
S^\pm_{J} &=&  \int d^dx dz \, \sqrt{g} \, e^{\pm\varphi(z)} \,
g^{K_1N_1} \cdots g^{K_{J-1/2}N_{J-1/2}} \,
\biggl[ \frac{i}{2} \bar\Psi^\pm_{K_1 \cdots K_{J-1/2}}
\epsilon_a^M \Gamma^a {\cal D}_M \Psi^\pm_{N_1 \cdots N_{J-1/2}}
\nonumber\\
&-& \frac{i}{2}
({\cal D}_M\Psi^\pm_{K_1 \cdots K_{J-1/2}})^\dagger
\Gamma^0 \epsilon_a^M \Gamma^a \Psi^\pm_{N_1 \cdots N_{J-1/2}}
- \bar\Psi^\pm_{K_1 \cdots K_{N-1/2}} \Big(\mu + V_F(z)\Big)
\Psi^\pm_{N_1 \cdots N_{J-1/2}} \biggr] \,. 
\en
As before, we remove the dilaton field from
the exponential prefactor, perform the boost of the spin-tensor field 
and restrict ourselves to the axial gauge. 
Next, after a straightforward algebra (including KK expansion),
we derive the same equation of motion for the KK profile and mass
formula as for fermions with lower spins. The action for
physical baryons with higher spins is written as
\eq
S_{J}^{(d)} = \sum\limits_n \, \int d^dx
\bar\Psi^{\mu_1 \cdots \mu_{J-1/2}, n}(x)
\biggl[ i \not\!\partial - M_{n} \biggr]
\Psi_{\mu_1 \cdots \mu_{J-1/2}}^n(x) \,.
\en
Therefore, the main difference
between the bosonic and fermionic actions is that in the case of bosons the
mass formula depends on the combination $(J+L)/2$, while in the baryon case
it depends only on $L$. Also in the fermion case the dilaton prefactor
and possible warping of conformal-invariant AdS metric can be easily
absorbed in the field, without the generation of extra potential terms.

\section{Conclusions}

We performed a systematic analysis of extra-dimensional actions for
bosons and fermions, which give rise to actions for observable hadrons. 
Masses are calculated analytically from Schr\"odinger type
equations of motion with a potential which provides confinement of the
Kaluza-Klein (KK) modes in extra $(d+1)$ dimension.
The tower of KK modes with radial quantum number $n$ and total
angular momentum $J$ has direct correspondence to realistic
mesons and baryons living in $d$ dimensions. For such correspondence
the sign of the dilaton profile is irrelevant, because the exponential
prefactor containing the dilaton is finally absorbed in the bulk fields.
On the other hand, the sign of the dilaton profile becomes important 
for the definition/calculation of the bulk-to-boundary propagator --
i.e. the Green function describing the evolution of bulk field from
inside of AdS space to its ultraviolet boundary. The corresponding
sign should be negative in order to fulfill certain constraints
discussed recently in Refs.~\cite{SWplus}.

\begin{acknowledgements}
The authors thank Stan Brodsky and Guy de T\'eramond for useful discussions
and remarks. 
V.E.L. would like to thank Departamento de F\'\i sica y Centro
Cient\'\i fico Tecnol\'ogico de Valpara\'\i so (CCTVal), Universidad
T\'ecnica Federico Santa Mar\'\i a, Valpara\'\i so, Chile for warm
hospitality.
\end{acknowledgements}

\end{document}